\documentstyle[preprint,aps,epsfig]{revtex}

\newcommand{\beq}{\begin{equation}}
\newcommand{\eeq}{\end{equation}}
\newcommand{\be}{\begin{eqnarray}}
\newcommand{\ee}{\end{eqnarray}}

\begin{document}

\draft
\preprint{
\begin{tabular}{l}
\hbox to\hsize{May, 2002\hfill KIAS-P02019}\\[-2mm]
\hbox to\hsize{              \hfill KAIST-TH 02/04}\\[-3mm]
\end{tabular}
}

\title{
Probing anomalous right-handed top quark couplings\\
in rare B decays
}

\author{
Kang Young Lee$^{1}$
\thanks{kylee@kias.re.kr},
Wan Young Song$^2$
\thanks{wysong@muon.kaist.ac.kr}
}

\address{
$^{1}$ School of Physics,
Korea Institute for Advanced Study, Seoul 130-012, Korea\\
$^{2}$ Department of Physics,
Korea Advanced Institute of Science and Technology,\\
Taejon 305-701, Korea\\
}

\date{\today}
\maketitle

\begin{abstract}
We explore 
the anomalous right-handed $\bar{t}sW$ and $\bar{t}bW$ couplings
using $ B \to X_s \gamma$ and $B \to X_s l^+ l^-$ decays
induced by the flavour-changing penguin diagrams.
The anomalous $\bar{t}sW$ coupling can yield 
10\% enhancement of the $B \to X_s l^+ l^-$ decay rate
under the constraint from the present $B \to X_s \gamma$ data,
while it does not affect the forward-backward asymmetry of the charged lepton.
The allowed region for anomalous $\bar{t}bW$ coupling
by the $ B \to X_s \gamma$ constraint is two-fold,
the small value region and the large value region.
Though the effect of the small anomalous coupling on the $B \to X_s l^+ l^-$ 
branching ratio is very little, 
it can yield a substantial chanhge for the forward-backward asymmetry.

\end{abstract}

\pacs{PACS numbers:12.90.+b,13.20.He,14.65.Ha}

\tightenlines


\section{Introduction}

After the discovery of the top quark at Tevatron\cite{top1,top2},
several properties of the top quark have been examined
such as top quark mass \cite{topmass},
production cross section \cite{topcross},
and the production kinematics \cite{topkin} etc..
The production of $10^7 - 10^8$ top quark pairs per year
is expected at Large Hadron Collider (LHC), 
which will allow us to study the detailed structure
of top quark couplings \cite{lhc}.
The top quark dominantly decays through the $t \to b W$ channel
and other channels are highly suppressed by small mixing angles.
Thus the  $\bar{t}bW$ coupling will be measured at LHC 
with high precision.
Effects of the anomalous $\bar{t}bW$ coupling have been studied 
in many literatures in direct and indirect ways 
\cite{larios,boos,rindani,elhady,yue}.
The subdominant channel in the Standard Model (SM) is
the Cabbibo-Kobayashi-Maskawa (CKM) nondiagonal decay $t \to s W$
of which branching ratio is estimated as
\be
Br(t \to s W) \sim 1.6 \times 10^{-3},
\ee
when $|V_{ts}| = 0.04$ is assumed.
Though the branching ratio of this channel is rather small,
the large number of expected top quark production at LHC 
will give us a chance to measure the $t \to s W$ process
and enable us to probe the $\bar{t}sW$ coupling 
responsible for this channel directly.
Therefore the anomalous $\bar{t}sW$ coupling,
which has not been seriously examined yet, is worth examining at present.

Before the LHC, we can study the top quark couplings indirectly 
in rare $B$ decays.
Rare $B$ decays involving loop induced flavour-changing neutral transitions
are sensitive to the properties of internal heavy particles,
so they can provide a good probe of new physics beyond the SM.
The radiative $b \to s \gamma$ and semileptonic $b \to s l^+ l^-$
decay are the most promising channels to examine the new physics effects.
The branching ratio of inclusive $B \to X_s \gamma$ decay 
has been measured by CLEO \cite{cleo}, ALEPH \cite{aleph} groups 
and recently by Belle collaboration 
from a 5.8 fb$^{-1}$ data sample \cite{belle1}.
We have the weighted average of the branching ratio of this channel as
\be
Br(B \to X_s \gamma) = (3.23 \pm 0.41 ) \times 10^{-4},
\ee
from those measurements.
This channel has intensively studied at next-leading order (NLO)
in the SM \cite{kagan,bsgamma}
and has provided stringent constraints on
various new physics models \cite{larios,susy,kim,cho,agashe}. 
On the other hand, the first observation of $b \to s l^+ l^-$ decay
is reported by the Belle group through
the exclusive $B\to K l^+ l^-$ channel \cite{belle2}
\be
Br(B \to K l^- l^+) = (0.75^{+0.25}_{-0.21}\pm 0.09) \times 10^{-6},
\ee
from a 30 fb$^{-1}$ data.
BaBar collaboration also present a bound on this mode
and the $B \to K^* \mu^- \mu^+$ mode \cite{babar}.
The inclusive  $B \to X_s l^+ l^-$ decay rate
is to be measured soon
as more data of $B$ decay will be accumulated.
The measurement of this mode provides 
a complementary study on the flavour-changing penguin decays.

In this work, we examine the effects of 
the anomalous right-handed $\bar{t}bW$ and $\bar{t}sW$ couplings 
on the inclusive $B \to X_s \gamma$ and $B \to X_s l^+ l^-$ decays.
We concentrate on the anomalous couplings of charged current interactions
and ignore effects of new particles andj the neutral current interactions.
With the anomalous right-handed couplings,
we write the effective lagrangian as
\be
{\cal L} = -\frac{g}{\sqrt{2}} 
          \sum_{q=s,b} V_{tq}~ \bar{q} \gamma^\mu 
                                 (P_L + \xi_q P_R) t W^-_\mu
  + H.c.,
\ee
where $\xi_q$ measures the new physics effects.
In section 2, we present the effective Hamiltonian approach
with the effective lagrangian Eq. (4).
The $B \to X_s \gamma$ and $B \to X_s l^+ l^-$ decays are
described in terms of the effective Hamiltonian
and the effects of the anomalous couplings are analyzed 
in section 3.
Our conclusion is given in section 4.

\section{The effective Hamiltonian}

In order to study the rare decay processes of $B$ meson,
the effective field theoretical approach is required to incorporate
the consistent QCD correction, which is substantial in rare
$B$ decays.
We can write the $\Delta B =1$ effective Hamiltonian to describe
$b \to s \gamma$ and $b \to s l^+ l^-$ processes as
\be
{\cal H}_{eff} = -\frac{4 G_F}{\sqrt{2}} V_{ts}^* V_{tb}
           \sum_{i=1}^{10}
             \left( C_i(\mu) O_i(\mu) + C'_i(\mu) O'_i(\mu) \right),
\ee
where the dimension 6 operators $O_i$ constructed in the SM 
are given in the Ref. \cite{buras},
and $O'_i$ are their chiral conjugate operators.
Matching the effective theory (5) and the lagrangian (4)
at $\mu = m_W$ scale,
we have the Wilson coefficients $C_i(\mu=m_W)$ and $C'_i(\mu=m_W)$.
When we let $\xi_q =0$, we have the Wilson coefficients in the SM 
\be
C_2(m_W) &=& -1,
\nonumber \\
C_7(m_W) &=& F(x_t),
\nonumber \\
C_8(m_W) &=& G(x_t),
\nonumber \\
C_9(m_W) 
&=& C_9^{\gamma} + C_9^Z + C_9^{\Box}
\nonumber \\
    &=& 
-D_0(x_t) 
-4 \left(1-\frac{1}{4 \sin^2 \theta_{W}} \right) C_0(x_t)
-\frac{1}{\sin^2 \theta_{W}} B_0(x_t),
\nonumber \\
C_{10}(m_W) 
&=& C_{10}^Z + C_{10}^{\Box}
\nonumber \\
         &=& -\frac{1}{\sin^2 \theta_{W}} C_0(x_t)
                     +\frac{1}{\sin^2 \theta_{W}} B_0(x_t),
\nonumber \\
C_i(m_W) &=& C'_i(m_W) = 0, ~~~~~~{\rm otherwise,}
\ee
where $F(x), G(x), D_0(x), C_0(x), B_0(x)$ are the well-known 
Inami-Lim loop functions \cite{buras,inami}:
\be
F(x) &=& \frac{x(7-5x-8x^2)}{24(x-1)^3}
                 - \frac{x^2 (2-3x)}{4(x-1)^4} \ln x,
\nonumber \\
G(x) &=& \frac{x(2+5x-x^2)}{8(x-1)^3}
                 - \frac{3x^2}{4(x-1)^4} \ln x,
\nonumber \\
D_0(x) &=& -\frac{4}{9} \ln x
                 + \frac{x^2(25-19x)}{36(x-1)^3}
        + \frac{x^2(5x^2-2x-6)}{18(x-1)^4} \ln x,
\nonumber \\
C_0(x) &=& \frac{x}{8} \left(
             \frac{x-6}{x-1} + \frac{3x+2}{(x-1)^2} \ln x \right),
\nonumber \\
B_0(x) &=& \frac{1}{4} \left(
             \frac{x}{1-x} + \frac{x}{(x-1)^2} \ln x \right).
\ee
Let us switch on the right-handed $\bar{t} b W$
and $\bar{t} s W$ couplings.
Keeping the effects of anomalous couplings in linear order,
we obtain the modified Wilson coefficients
\be
C_7 &\to& C_7^{{\rm SM}} + \xi_b \frac{m_t}{m_b} \tilde{F}(x_t),
\nonumber \\
C_8 &\to& C_8^{{\rm SM}} + \xi_b \frac{m_t}{m_b} \tilde{G}(x_t),
\nonumber \\
C_9 &\to& C_9^{{\rm SM}} - \xi_b \frac{m_b}{m_t} \tilde{D}(x_t),
\ee
and the new Wilson coefficients
\be
C'_7 &=& \xi_s \frac{m_t}{m_b} \tilde{F}(x_t),
\nonumber \\
C'_8 &=& \xi_s \frac{m_t}{m_b} \tilde{G}(x_t),
\nonumber \\
C'_9 &=& -\xi_s \frac{m_b}{m_t} \tilde{D}(x_t),
\ee
with the new loop functions
\be
\tilde{F}(x) &=& \frac{-20+31x-5x^2}{12(x-1)^2}
                 + \frac{x (2-3x)}{2(x-1)^3} \ln x,
\nonumber \\
\tilde{G}(x) &=& -\frac{4+x+x^2}{4(x-1)^2}
                 + \frac{3x}{2(x-1)^3} \ln x,
\\
\tilde{D}(x) &=& \frac{x(59-38x+25x^2+2x^3)}{36(x-1)^4}
            - \frac{2(x+1)}{3(x-1)^5} \ln x
                 - \frac{x^2}{2(x-1)^4} \ln x.
\nonumber 
\ee
Our new loop functions $\tilde{F}(x)$ and $\tilde{G}(x)$
agree with those in Ref. \cite{cho} and 
$\tilde{D}(x)$ is the first calculation.
Note that the ${\cal O}(\xi)$ terms of $Z$--penguin diagram are
suppressed by the heavy mass of $Z$--boson as $m_b^2/m_Z^2$,
or $q^2/m_Z^2$ and we neglect them here.
For the box diagram, the ${\cal O}(\xi)$ terms vanish by the chirality
relation and the leading contribution is of $\xi^2$ order.
As a consequence, the contribution of order ${\cal O}(\xi)$ 
comes only through the $\gamma$-penguin and gluon penguin diagrams.
Thus there exists no new effect in $C_{10}$ and $C'_{10}=0$.

The renormalization group (RG) evolution
of the Wilson coefficients ${\bf C} = (C_i,C'_i)^\dagger$ given by
\be
\mu \frac{d}{d \mu} {\bf C}(M_W)
  = -\frac{g^2}{16 \pi^2} \gamma^T {\bf C}(M_W),
\nonumber
\ee
is governed by a $20 \times 20$ anomalous dimension matrix $\gamma$.
Since the strong interaction preserves chirality,
the operators $Q'_i$ are evolved separately
without mixing between those and the SM operators. 
Thus the $20 \times 20$ anomalous dimension matrix $\gamma$ is decomposed 
into two identical $10 \times 10$ matrices $\gamma_{0}$ given in the SM 
as
\be
\gamma = \left( \begin{array}{c}
             \gamma_{0}~~~~~0~ \\
             ~~0~~~~~\gamma_{0}~
           \end{array}   \right) ,
\nonumber
\ee
The $10 \times 10$ anomalous dimension matrix $\gamma_0$ 
has been calculated to leading logarithmic level 
in Ref. \cite{ciuchini,buras0}.
Using the initial condition at $\mu = m_W$,
\be
(C_i(M_W), C'_i(M_W))
= (0,-1,0,0,0,0,C_7,C_8,C_9,C_{10},0,0,0,0,0,0,C'_7,C'_8,C'_9,0),
\ee
we can solve the RG equation to obtain the Wilson coefficients
evolved from $\mu = m_{W}$ to $\mu = m_b$ scales. 

\section{Rare $B$ decays}

\subsection{$B \to X_s \gamma$}

The branching ratio of $B \to X_s \gamma$ process 
with the right-handed interactions are obtained at NLO 
\be
{\rm Br}(B \to X_s \gamma) &=& \frac{{\rm Br}(B \to X_c e \bar{\nu})}
                                    {10.5 \%}
\nonumber \\
&&~~~~\times \left[ B_{22}(\delta) + B_{77}(\delta) (|r_7|^2 + |r'_7|^2 )
+ B_{88}(\delta) (|r_8|^2 + |r'_8|^2 )
\right.
\nonumber \\
&&~~~~~~ \left.+ B_{27}(\delta) Re (r_7)
+ B_{28}(\delta) Re (r_8)
+ B_{78}(\delta) ( Re (r_7 r^{\star}_8) + Re (r'_7 r'^{\star}_8) )
\right],
\ee
where the ratios $r_i$ and $r'_i$ are defined by
\be
r_i = \frac{C_i(m_W)}{C_i^{SM}(m_W)} 
    = 1 + \frac{C_i^{\rm New}(m_W)}{C_i^{SM}(m_W)},
~~~~~~~~
r'_i = \frac{C'_i(m_W)}{C_i^{SM}(m_W)}.
\ee
The components $B_{ij}(\delta)$ depends on the kinematic cut $\delta$,
of which numerical values are given in the Ref. \cite{kagan}.

With the measured branching ratio, Eq. (2), 
we can set the conservative bounds on the parameter 
$\xi_{b}$ and $\xi_{s}$ as
\be
-0.0021 < \xi_b < 0.0031~~~&& {\rm (A)},~~~~~~ 
-0.0485 < \xi_b < -0.0433~~~ {\rm (B)},
\nonumber \\
&&| \xi_s| < 0.012,
\ee
at 2-$\sigma$ level.
We assume that $\xi_b$ and $\xi_s$ are real for simplicity.
The anomalous coupling $\xi_b$ contributes 
in the linear and quadratic order
while $\xi_s$ dominantly contributes in the quadratic order 
since the contribution of the linear order is strongly
suppressed by the ratio $m_s^2/m_b^2$. 
Thus the parameter $\xi_s$ is less constrained by the $B \to X_s \gamma$
measurement than $\xi_b$ in general.
Due to the cancellation by the interference term $B_{27}~Re(r_7)$,
however,
the large $|\xi_b|$ solution in the region B is also allowed,
which gives the positive Wilson coefficient $C_7(m_W) > 0$.

\subsection{$B \to X_s l^+ l^-$}

The dilepton invariant mass distribution of $B \to X_s l^+ l^-$ decays
consists of following contributions :
\be
\frac{d {\rm Br} (B \to X_s l^- l^+)}{d \hat{s}} 
= \frac{dB_0}{d \hat{s}}
+ \frac{dB_{1/m_b^2}}{d \hat{s}}
+ \frac{dB_{1/q^2}}{d \hat{s}},
\ee
where the first term denotes the decay at the parton level,
the second term the power correction 
in the heavy quark effective theory (HQET),
and the last term is due to the nonperturbative virtual quark loop effects
with soft gluon.
We have the explicit expression including the HQET corrections 
of order ${\cal O}(1/m_b^2)$ given in Ref. \cite{ali2,ali1,munz},
\be
\frac{d {\rm Br} (B \to X_s l^- l^+)}{d \hat{s}} &=& 2 B_0
\left[
\left( \frac{1}{3} (1-\hat{s})^2 (1+2 \hat{s}) (2+\hat{\lambda}_1)
      +(1 - 15 \hat{s}^2 + 10 \hat{s}^3 ) \hat{\lambda}_2) \right)
\right.
\nonumber \\
&&~~~~~~~\times (|C_9|^2 + |C'_9|^2 + |C_{10}|^2)
\nonumber \\
&&~~~ \left. + \left(\frac{4}{3} (1-\hat{s})^2 (2+\hat{s}) (2+\hat{\lambda}_1)
      +4 (-6-3\hat{s} + 5 \hat{s}^3 ) \hat{\lambda}_2) \right)
    \frac{(|C_7|^2 + |C'_7|^2 )}{\hat{s}}
\right.
\nonumber \\
&&~~~ \left. + \left(4 (1-\hat{s})^2 (2+\hat{\lambda}_1)
      +4 (-5-6\hat{s} + 7 \hat{s}^2 ) \hat{\lambda}_2) \right)
     {\rm Re}(C_7 C_9^* +C'_7 {C'}_9^*)
\right],
\ee
where $\hat{s} = (p_+ + p_-)^2/m_b^2$ is
the normalized dilepton invariant mass, 
$\hat{\lambda}_{1,2} = \lambda_{1,2}/m_b^2$ 
the normalized HQET parameters, 
and the normalization constant is given by
\be
B_0 \equiv {\rm Br}(B \to X_c e \bar{\nu})
              \frac{3 \alpha^2}{16\pi^2}
              \frac{|V_{ts}^* V_{tb}|^2}{|V_{cb}|^2}
              \frac{1}{f(\hat{m}_c) \kappa(\hat{m}_c)},
\ee
with the phase space function $f(\hat{m}_c)$ 
and the perturbative QCD correction $\kappa(\hat{m}_c)$
of $B \to X_c l \nu$ decay.
The long-distance correction due to the virtual $c \bar{c}$ loop
is considered in Ref. \cite{buchalla}.
Numerically this correction is at 1 - 2\% level 
in the region of $\hat{s}$ considered here away from the resonances
and we ignore them.
We plot the differential branching ratio with respect to
$\hat{s} $ in the Fig. 1. 
The values of $\xi_{b,s}$ given in Eq. (14) are used.
We consider the region for $\hat{s}$ as 
1 GeV$^2 < \hat{s} m_b^2 <$ 7.5 GeV$^2$ in order to avoid 
the large resonant contribution of $J/\psi$ and $\psi'$.
We show that the total decay rate over this region is enhanced
by the anomalous couplings.
The $\xi_s = 0.012$ leads to 10\% enhancement of the branching ratio.
The enhancement by $\xi_b$ in the region A is less than 1\%
and the branching ratio with $\xi_b$ in the region B
is more than twice the SM prediction since there is
no interference term between $C_2$ and $C_7$
in the $B \to X_s l^- l^+$ decay rate,
which cancels the enhanced $|C_7|^2$ contribution. 
Hence such a enhancement should be observed in the near future,
if $\xi_b$ coupling in the region B exists.

The forward-backward (FB) asymmetry $A_{FB}$ is defined as
\be
\frac{dA_{FB}}{d \hat{s}} = \int_0^1 
            \frac{ d^2 {\rm Br}}{d \hat{s}~d \cos \theta} d \cos \theta
                     - \int_{-1}^0 
            \frac{ d^2 {\rm Br}}{d \hat{s}~d \cos \theta} d \cos \theta,
\ee
where the angle $\theta$ is measured between $b$-quark
and the positively charged lepton $l^+$ 
in the dilepton center-of-mass (CM) frame.
The $A_{FB}$ distribution with respect to $\hat{s}$
is depicted in the Fig. 2.
We find that $\xi_b$ in the region A can bring a substantial shift 
of the differential FB asymmetry
although the branching ratio is not much affected.
It is because the branching ratio is dominated by the SM contribution
$|C_9|^2 + |C_{10}|^2$ but $A_{FB}$ is proportional
to the substantial Wilson coefficient $C_{10}$ and shifted
by the term $\sim \xi_b C_{10}$.
As a consequence, $A_{FB}$ summed over the region considered here
is enhanced by 4 times with $\xi_b = -0.0021$
and even its sign is changed if $\xi_b = 0.0031$.
With $\xi_b$ in the region B, the shift of $dA_{FB}/d \hat{s}$
is huge as is the case of decay rate, because of the large shift of $C_7$.
Besides, the anomalous $\bar{t} s W$ coupling does not affect $A_{FB}$,
since $C'_{10}=0$.

\section{Concluding Remarks}

We studied rare $B$ decays with the anomalous right-handed
$\bar{t}bW$ and $\bar{t}sW$ couplings in the model-independent way.
This kind of anomalous couplings can be obtained 
in the general left-right (LR) model
based on SU(2)$_L \times$SU(2)$_R \times$U(1) gauge group \cite{LR}
or the dynamical electroweak symmetry breaking model \cite{zhang}.
In the LR model, the right-handed quark mixing is also an observable.
If we do not demand the symmetry 
between left- and right-handed sectors manifest, 
the right-handed quark mixing is not necessarily identical
to the left-handed quark mixing described by the CKM matrix.
Thus we have right-handed charged current interactions,
which is suppressed by the heavy mass of extra $W$ boson.
but still enhanced relatively
by the right-handed quark mixing matrix.
On the other hand, when the electroweak symmetry breakdown is dynamical,
one may expect that some nonuniversal interactions exist
which lead to anomalous couplings on charged current interactions.

The constrainyt on $\xi_s$ by the $B \to X_s \gamma$ data 
is weaker than that of $\xi_b$ in the region A
by the chirality relation.
As a result, the considerable enhancement of the branching ratios
for $B \to X_s l^+ l^-$ decay is possible with 
the anomalous $\bar{t}sW$ coupling 
while the influence of the anomalous $\bar{t}bW$ coupling 
in the region A is rather small.
Besides anomalous $\bar{t}bW$ coupling
can change the forward-backward asymmetry considerably
under the $B \to X_s \gamma$ constraints,
and even the sign of $A_{FB}$ may be reversed
while the anomalous $\bar{t}sW$ coupling does not affect $A_{FB}$.
Therefore it is possible to discriminate the effects of
the anomalous $\bar{t}bW$ and $\bar{t}sW$ couplings
if we combine the analysis of the branching ratio and $A_{FB}$ 
for $B \to X_s l^+ l^-$ process.
On the other hand, relatively large value of $|\xi_b|$ in the region B 
is also allowed due to the cancellation 
between the $C_2 C_7$ and $|C_7|^2$ terms
for $B \to X_s \gamma$ decay,
which leads to much larger branching ratio and altered $A_{FB}$ 
for $B \to X_s l^+ l^-$ decay.
The measured branching ratio of the exclusive 
$B \to K l^+ l^-$ decay given in Eq. (3), 
is rather higher than the SM prediction,
though it is still consistent with the SM \cite{exclusive} 
due to large errors and theoretical uncertainties.
Thus it may be the clue of the new physics signal
and the anomalous coupling $\xi_b$ in the region B
can be a candidate of the new physics.
If we measure $A_{FB}$ in the future, it will be a clear probe
of the nature of the top quark anmalous couplings.

\acknowledgments
We thank T. Morozumi and T. Goto for valuable comments
and K. Hagiwara for hospitality.
This work is supported in part by BK21 Core program of the Ministry of
Education (MOE).

\begin{figure}[htb]
\begin{center}
\epsfig{file=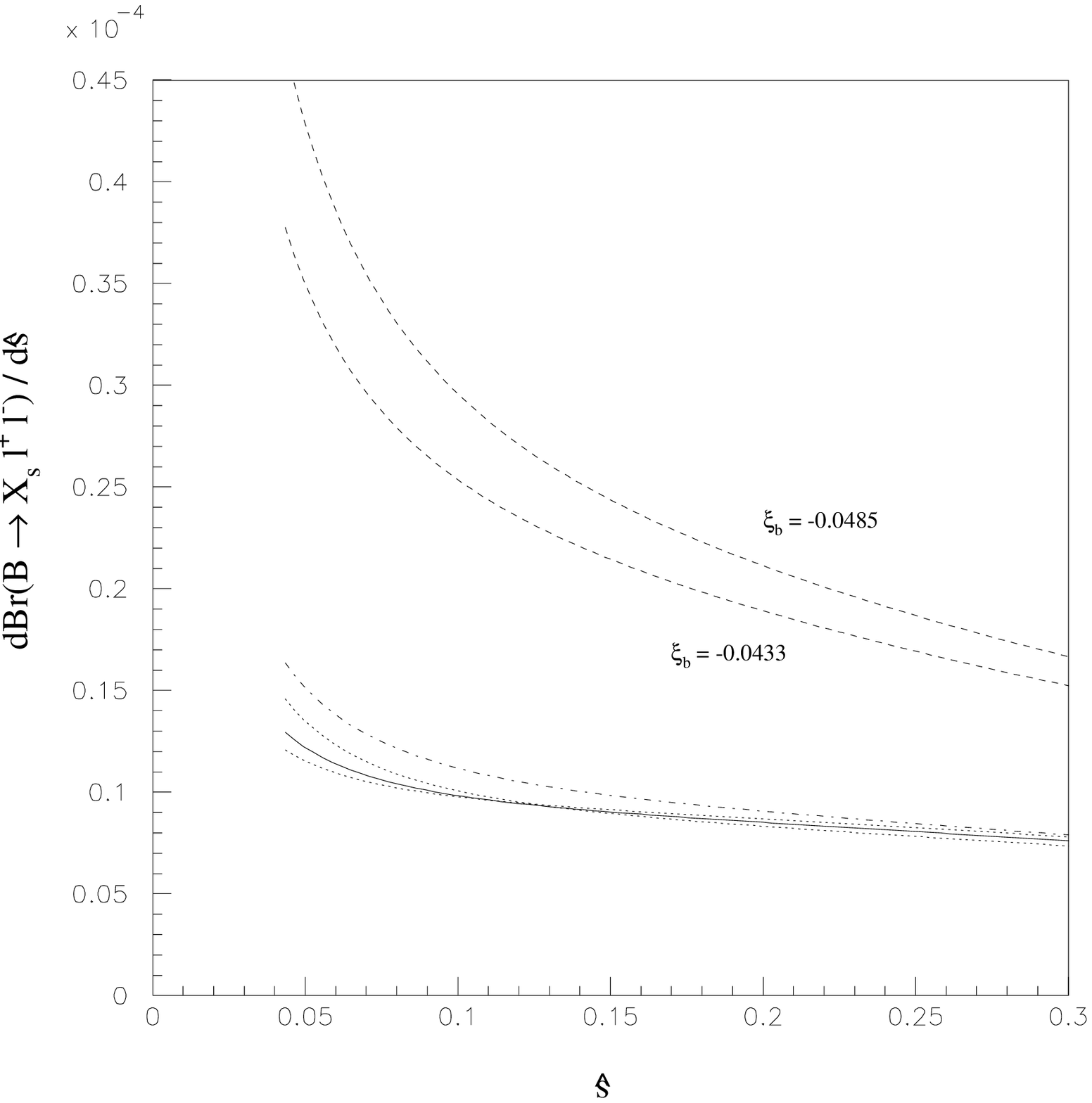,height=18cm}
\caption{
The differential branching ratio of $B \to X_s l^+ l^-$ decays.
The solid line denotes the SM prediction, 
the dotted line the prediction with $\xi_b$ in the region A,
the dashed line the prediction with $\xi_b$ in the region B,
and the dash-dotted line the prediction with $\xi_s = 0.012$.
}
\label{fig1}
\end{center}
\end{figure}

\begin{figure}[htb]
\begin{center}
\epsfig{file=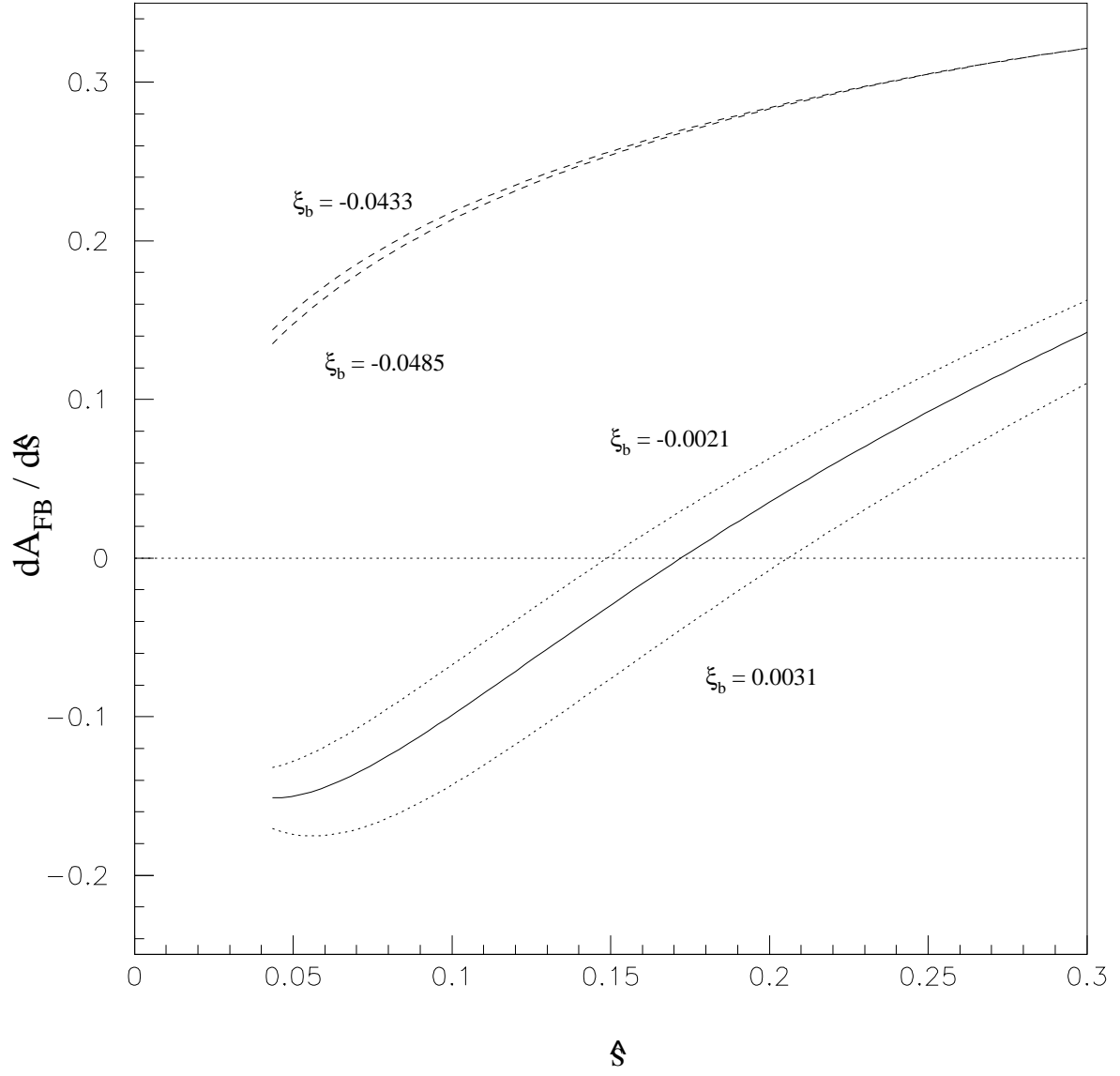,height=18cm}
\caption{
The forward-backward asymmetry of $B \to X_s l^+ l^-$ decays.
The solid line denotes the SM prediction, 
the dotted line the prediction with $\xi_b$ in the region A,
the dashed line the prediction with $\xi_b$ in the region B.
}
\label{fig2}
\end{center}
\end{figure}

\end{document}